\documentclass{article}

    \usepackage[preprint, nonatbib]{neurips_2021}

\usepackage[utf8]{inputenc} 
\usepackage[T1]{fontenc}    
\usepackage{hyperref}       
\usepackage{url}            
\usepackage{booktabs}       
\usepackage{amsfonts}       
\usepackage{nicefrac}       
\usepackage{microtype}      
\usepackage{xcolor}         
\usepackage[pdftex]{graphicx}
\title{Evaluating robustness of You Only Hear Once (YOHO) Algorithm on noisy audios in the VOICe Dataset}

\author{%
  Soham Tiwari \\
  Department of Computer Science \& Engineering\\
  Manipal Institute of Technology, Manipal\\
  \texttt{soham.tiwari@learner.manipal.edu} \\
   \And
   Kshitiz Lakhotia \\
   Department of Information and Communication Technology\\
   Manipal Institute of Technology, Manipal\\
  \texttt{kshitiz.lakhotia1@learner.manipal.edu} \\

   \AND
   Manjunath Mulimani \\
   Department of Computer Science \& Engineering \\
   Manipal Institute of Technology, Manipal \\
   \texttt{manjunath.mulimani@manipal.edu}
}

\begin{document}

\maketitle

\begin{abstract}
Sound event detection (SED) in machine listening entails identifying the different sounds in an audio file and identifying the start and end time of a particular sound event in the audio. SED finds use in various applications such as audio surveillance \cite{1_audio-surveillance}, speech recognition, and context-based indexing and retrieval of data in a multimedia database \cite{2_sound-event-detection}. However, in real-life scenarios, the audios from various sources are seldom devoid of any interfering noise or disturbance. In this paper, we test the performance of the You Only Hear Once (YOHO) \cite{3_yoho} algorithm on noisy audio data. Inspired by the You Only Look Once (YOLO) \cite{7_yolo} algorithm in computer vision, the YOHO algorithm can match the performance of the various state-of-the-art algorithms on datasets such as Music Speech Detection Dataset \cite{5_music_speech}, TUT Sound Event  \cite{6_TUT}, and Urban-SED datasets \cite{25_urban-sed} but at lower inference times. In this paper, we explore the performance of the YOHO algorithm on the VOICe dataset \cite{4_voice} containing audio files with noise at different sound-to-noise ratios (SNR). YOHO could outperform or at least match the best performing SED algorithms reported in the VOICe dataset paper and made inferences in less time.
\end{abstract}

\section{Introduction}

Sound event detection\footnote{\href{https://github.com/sohamtiwari3120/YOHO-on-VOICe}{GitHub Code}} (SED) involves detecting the type as well as
the onset and offset times of sound events in audio streams, i.e., the task of recognizing the sound events and their respective temporal start and end time in a recording \cite{8_sound_tutor}. As a result, SED can be utilized in a variety of applications. These include context-based indexing and retrieval in multimedia databases, bio-acoustic monitoring, intelligent remixing, and surveillance. Moreover, the detected events are used as intermediate representations in other research areas/applications, e.g., audio context recognition, automatic tagging, and audio segmentation \cite{2_sound-event-detection}. However, building SED models is difficult because sound events exhibit diverse temporal and spectral characteristics, and because they rarely occur in isolation, multiple sound events tend to overlap considerably (polyphonic audio).  

A multitude of models has been used to detect sound events over the years. Early approaches for sound event detection had adapted techniques from speech recognition and music information retrieval, such as Gaussian Mixture Models (GMMs) and Hidden Markov Models (HMMs) \cite{8_sound_tutor}. However, a drawback of HMMs is the inability to deal with polyphonic audio - containing multiple sound events in parallel. In contrast, the advent of deep learning has shown that deep neural networks are adept at performing multi-label sound event classification and dealing with polyphonic audio. A simple feed-forward neural network (FFN) can perform multi-label classification but is limited in its ability to model the long-term time correlation of the sound signals.
Consequently, plain FNNs are not widely used in SED. While convolutional neural networks (CNNs) are not suitable for modeling the time correlations of audio signals alone, CNNs combined with recurrent neural networks (RNNs) can capture the local and global context in audios. Convolutional Recurrent neural networks (CRNNs) \cite{9_CRNN} perform exceptionally well in SED. The convolutional layers act as feature extractors, which learn discriminatory features through the successive convolutions and non-linear transformations applied to the time-frequency spectrogram representation presented at the network's input. The recurrent layers learn the temporal dependencies in the sequence of features presented at their input. However, sequential layers preclude full use of parallel computation in GPUs, leading to longer times for performing sound event detection on audio files.
As a result, to reduce inference times without sacrificing high precision and recall, S. Venkatesh et al. proposed the You Only Hear Once (YOHO) algorithm. YOHO algorithm was inspired by You Only Look Once (YOLO) \cite{7_yolo} algorithm, which has tremendous popularity in Computer Vision. YOLO converted the prediction of object bounding boxes from a classification problem to a regression problem, drastically improving the input image processing times and facilitating real-time object detection. Similarly, YOHO converts the detection of acoustic boundaries from a frame-based, multi-label classification problem into a regression problem. This was achieved by having a set of 3 separate output neurons for each class. One neuron to detect the presence of an event class and the other two to predict the event's start and endpoints. The YOLO paper \cite{3_yoho} reported a higher F-measure and lower error rate than CRNNs on multiple datasets.
Moreover, since YOHO is designed as a purely convolutional neural network with no recurrent layers, it has faster inference times than architectures that include sequential layers like CRNNs. It takes as input log-Mel spectrograms \cite{10_logmel} of the audios and outputs the probability as well as the start and end time of each event label. In addition, as this end-to-end approach predicts acoustic boundaries directly, it is significantly quicker during post-processing and smoothing \cite{3_yoho}.

To test the performance of YOHO on noisy audio, we used the VOICe dataset \cite{4_voice}. It is an artificially created dataset using sound events from TUT Rare Sound Events 2017 development set and acoustic scenes from TAU Urban Acoustic Scenes 2019 development set, containing different sound events combined with noise from different acoustic scenes. The motivation behind this work is to test and evaluate the performance of YOHO on noisy data. If it can perform well on noisy data and report scores matching the CRNN algorithms, we would then have in our hands an algorithm that would be robust to noise and be able to perform sound event detection much faster than other algorithms today. As a result, it would then turn out to be a candidate model deployed on portable devices that want to carry out sound event detection.
 The rest of the paper is structured as follows. Section 2 explores the VOICe dataset, followed by Section 3, where the YOHO model adapted for the VOICe dataset is described. In section 4, the SED results on VOICe are discussed, and finally, in section 5, the conclusion.

\section{VOICe Dataset}
\label{gen_inst}

The performance of sound event detection algorithms suffer when utilized in unknown situations, for example in different acoustic environments. Hence, the VOICe dataset \cite{4_voice} is used to develop and test sound event algorithms in different domains.
 
Audio noise from three acoustic environments: "vehicle", "outdoor", and "indoor" were utilized to simulate a realistic scenario in which the acoustic scene circumstances in the audio recordings of data for which a model is optimized and data for which the model is evaluated on \cite{4_voice} is incongruent. 

To make the sound event mixtures, isolated audio samples from three different sound event classes were used - "baby crying", "glass breaking" and "gunshot". Apart from retaining the clean audio files in the dataset, the clean audios were then augmented using background noise from recordings of cars, outdoor environments, and indoor acoustic scenes, yielding four separate versions of the same number of sound event mixtures at two different signal-to-noise ratios (SNRs) \cite{15_snr}: -3dB and -9dB, respectively.

As a result, VOICe consisted of 1449 different mixtures of three different sound events: "baby crying," "glass breaking," and "gunshot" in different conditions. 1242 mixtures with background noise of three different categories of acoustic scenes - "vehicle"," outdoor", and "indoor", were mixed using 2 SNR values (-3, -9 dB), i.e., 207 mixtures x 3 acoustic scenes x 2 SNRs = 1242. Moreover 207 clean audios without any background noise were included as well.

These audio files were then segregated into 'training', 'testing' and 'validation' splits. The training audio files were approximately 180s long where as testing and validation files were both 60s long.

Consequently, VOICe was designed to facilitate the change of the domain of sound event detection from one acoustic scene to another or between sound events with and without background noise.

\section{YOHO Algorithm}
 
S. Venkatesh et al. implemented the YOHO algorithm \cite{3_yoho} by altering the final layers of MobileNet \cite{13_mobilenet} architecture instead of the traditional, larger YOLO architectures \cite{7_yolo, 12_yolo9000}. YamNet \cite{14_yamnet}, although a simple audio classifier, is another algorithm that used MobileNet.

As a result, YOHO is a purely convolutional neural network, as seen in Table \ref{tab:tableone}. The upper half of the table consists of the original MobileNet \cite{13_mobilenet} architectural layers. The lower half contains layers modified for the VOICe dataset \cite{4_voice}. 

The input to the YOHO model is in the form of log-mel spectrograms \cite{16_logmel}. Each audio file is first segmented into overlapping windows with window length 2.56s and hop length 1.96s. The input dimension is determined by the length of the audio example and the number of mel bins. For the VOICe dataset, the input logmel spectrogram shape is (40, 257) where 40 is the number of mel bins and 257 time steps. Next, after reshaping the input, a 3x3 2D convolution with a stride of 2 is performed. As a result, the time and frequency dimensions are cut in half. Many depthwise-separable convolutions \cite{17_depthwise} with 3x3 filters are used in the MobileNet design, followed by pointwise convolutions \cite{18_pointwise} with 1x1 filters. Except for the final layer, all convolutions were followed by ReLu activations \cite{20_relu} and batch normalization \cite{19_batchnorm}. The time and frequency dimensions are halved each time a stride of two is used.

\begin{figure}[ht]
\caption{Representation of the output layer of YOHO for the VOICe dataset.}
\centering
\includegraphics[width=1\linewidth]{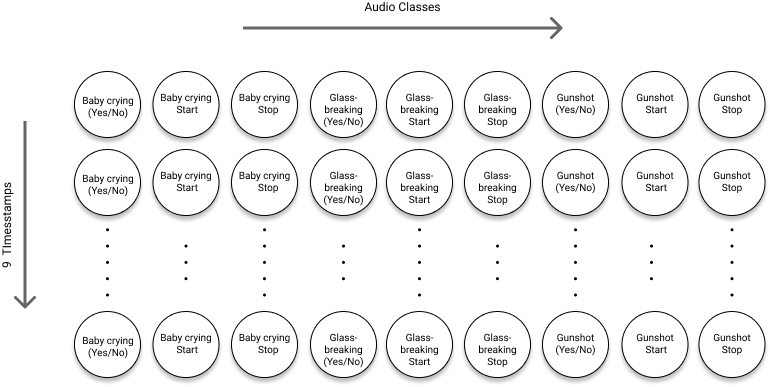}
\label{fig:output-layer}
\end{figure}

The last two dimensions were flattened. The final nine-filter 1D convolution yields an output matrix of shape 9x9. The first dimension of the output corresponds to the time axis and the second dimensions corresponds to 3 neurons for each of the 3 classes in the VOICe dataset. The input spectrogram is divided into 9 bins along the time axis. Every row in the output matrix represents the start and end times of the audio events in the corresponding bin of the input spectrogram. 
As seen in Figure \ref{fig:output-layer}, the first, fourth and seventh neurons perform binary classification at each time step to detect the presence of the respective audio events. The second and third neurons use regression to predict start and end times of the first audio-event class. Similarly for the fifth, sixth, eight and ninth  neurons and their respective second and third audio-event classes. The YOHO algorithm's output layer modified for the VOICe dataset is depicted in Figure 1.


\begin{table}[ht]

\centering
\caption{\label{tab:tableone}First half of the table consists of the original layers of the MobileNet \cite{11_mobilenet} architecture. The second half of the architecture has been adapted for the VOICe dataset. 'dw' stands for depth-wise CNN layer.}

\begin{tabular}{|c|c|c|c|}
\hline
{ \textbf{Layer type}}                                        & \textbf{Filters}                                   & {\textbf{Shape/Stride}}                            & { \textbf{Output shape}}                                            \\ \hline
Reshape                                                          & -                                                  & -                                                        & 257 x 40 x 1                                                           \\ \hline
Conv2D                                                           & 32                                                 & 3 x 3 / 2                                                & 129 x 20 x 32                                                          \\ \hline
Conv2D-dw                                                        & -                                                  & 3 x 3                                                    & 129 x 20 x 32                                                          \\ \hline
Conv2D                                                           & 64                                                 & 1 x 1                                                    & 129 x 20 x64                                                           \\ \hline
Conv2D-dw                                                        & -                                                  & 3 x 3 / 2                                                & 65 x 10 x 64                                                          \\ \hline
Conv2D                                                           & 128                                                & 1 x 1                                                    & 65 x 10 x 128                                                         \\ \hline
Conv2D-dw                                                        & -                                                  & 3 x 3                                                    & 65 x 10 x 128                                                         \\ \hline
Conv2D                                                           & 128                                                & 1 x 1                                                    & 65 x 10 x 128                                                         \\ \hline
Conv2D-dw                                                        & -                                                  & 3 x 3 / 2                                                & 33 x 5 x 128                                                          \\ \hline
Conv2D 256                                                       & 256                                                & 1 x 1                                                    & 33 x 5 x 256                                                          \\ \hline
Conv2D-dw                                                        & -                                                  & 3 x 3                                                    & 33 x 5 x 256                                                          \\ \hline
Conv2D 256                                                       & 256                                                & 1 x 1                                                    & 33 x 5 x 256                                                          \\ \hline
Conv2D-dw                                                        & -                                                  & 3 x 3 / 2                                                & 17 x 3 x 256                                                           \\ \hline
Conv2D                                                           & 512                                                & 1 x 1                                                    & 17 x 3 x 512                                                           \\ \hline
\begin{tabular}[c]{@{}c@{}} 5x ( Conv2D-dw\\ Conv2D)\end{tabular} & \begin{tabular}[c]{@{}c@{}}-\\ \\ 512\end{tabular} &
\begin{tabular}[c]{@{}c@{}}3 x 3\\ \\ 1 x 1\end{tabular} & \begin{tabular}[c]{@{}c@{}}17 x 3 x 512\\ \\ 17 x 3 x 512\end{tabular} \\ \hline
Conv2D-dw                                                        & -                                                  & 3 x 3 / 2                                                & 9 x 2 x 512                                                           \\ \hline
Conv2D                                                           & 1024                                               & 1 x 1                                                    & 9 x 2 x 1024                                                          \\ \hline
Conv2D-dw                                                        & -                                                  & 3 x 3                                                    & 9 x 2 x 1024                                                          \\ \hline
Conv2D                                                           & 1024                                               & 1 x 1                                                    & 9 x 2 x 1024                                                          \\ \hline 
\hline

\multicolumn{1}{|c|}{Conv2D-dw}  & \multicolumn{1}{c|}{-}         & \multicolumn{1}{c|}{3 x 3}            & \multicolumn{1}{c|}{9 x 2 x 1024}  \\ \hline
\multicolumn{1}{|c|}{Conv2D}      & \multicolumn{1}{c|}{512}        & \multicolumn{1}{c|}{1 x 1}            & \multicolumn{1}{c|}{9 x 2 x 512}   \\ \hline
\multicolumn{1}{|c|}{Conv2D-dw}   & \multicolumn{1}{c|}{-}         & \multicolumn{1}{c|}{3 x 3 / 2}        & \multicolumn{1}{c|}{9 x 2 x 512}   \\ \hline
\multicolumn{1}{|c|}{Conv2D}      & \multicolumn{1}{c|}{256}       & \multicolumn{1}{c|}{1 x 1}            & \multicolumn{1}{c|}{9 x 2 x 256}   \\ \hline
\multicolumn{1}{|c|}{Conv2D-dw}   & \multicolumn{1}{c|}{-}         & \multicolumn{1}{c|}{3 x 3}            & \multicolumn{1}{c|}{9 x 2 x 256}   \\ \hline
\multicolumn{1}{|c|}{Conv2D}      & \multicolumn{1}{c|}{128}       & \multicolumn{1}{c|}{1 x 1}            & \multicolumn{1}{c|}{9 x 2 x 128}   \\ \hline
\multicolumn{1}{|c|}{Reshape}     & \multicolumn{1}{c|}{-}         & \multicolumn{1}{c|}{-}                & \multicolumn{1}{c|}{9 x 256}       \\ \hline
\multicolumn{1}{|c|}{Conv1D}      & \multicolumn{1}{c|}{9}         & \multicolumn{1}{c|}{1}                & \multicolumn{1}{c|}{9 x 9}         \\ \hline
\end{tabular}
\end{table}

\section{Results}
\label{others}

Table \ref{tab:table-2} are the F1 scores  of the SED and adaptation tasks reported on the VOICe dataset using the CRNN algorithm \cite{9_CRNN} reported in the original paper. Here, ${D_s}$ and ${D_t}$ represent the source and the target domain data respectively. Furthermore, V. stands for "vehicle", O. is "outdoor", I. represents "indoor" and Ce. symbolises "clean" data.

The CRNN model in the original paper \cite{4_voice} was first trained on the source dataset $D_s$ - either clean audio files, vehicle audio files, indoor audio files or outdoor files using the provided train, test and validation splits. Next, a clone of this 'source' model - the 'target' model is made which initially contains the same weights. Moreover, a simple domain classifier consisting of two FNN layers was also made to classify the domain of the input data as either source or target. Finally the target model and the domain classifier were trained in an adversarial manner on the adaptation and the test splits of the target dataset $D_t$. The adaptation split is the train and validation splits of the target data combined \cite{4_voice}.

Hence, the paper \cite{4_voice} states that each row of Table \ref{tab:table-2} presents an adaptation process from $D_s$ to $D_t$. In addition, $F_{1_S}^{S_{te}}$ and $ F_{1_S}^{T_{te}}$ represents the F1-scores of source model for test split of $D_s$ and $D_t$ respectively. Conversely, $F_{1_A}^{T_{te}}$ denotes the F1-score of adapted model (i.e. target model) on test split data from $D_t$.

However, due to lack of powerful computational resources \footnote{Our models were trained on Google Colab using the free tier GPU.}, we were unable to use the adversarial method of training in our data. Hence, we do not have any results for a target YOHO model which would be trained on source data and adapted to target data in an adversarial manner. Instead, in our experiment the YOHO model would first be trained using the train and the validation splits of the source dataset $D_s$ and the same source model would be simply evaluated on the test split of the target dataset $D_t$.

Table \ref{tab:yoho9dB} contains the results of training the YOHO model on the source dataset $D_s$ and then simply evaluating on the target dataset $D_t$. This table only contains results for -9dB SNR data \footnote{The results for the -3dB SNR data using YOHO were not reported due to the lack of benchmarks for comparison on -3dB SNR data using CRNNs in the original VOICe paper \cite{4_voice}.}. Hence comparing the results in Table 3 with only the $F_{1_s}^{S_te}$ and $F_{1_s}^{T_te}$ columns of Table 2, the YOHO model reports higher F1-scores compared to the CRNN \cite{9_CRNN} model on VOICe in the following:
\begin{itemize}
    \item When the model is trained on clean data as source $D_s$, it has obtained higher F1-scores across all target datasets $D_t$ (except when $D_t$ is clean) compared to CRNNs.
    \item When the model is trained on vehicle data as source $D_s$, it has obtained higher F1-scores across all target datasets $D_t$.
    \item When the model is trained on outdoor data as source $D_s$, it has obtained higher F1-scores across all target datasets $D_t$.
    \item When the model is trained on indoor data as source $D_s$, it has obtained higher F1-scores across all target datasets $D_t$.
\end{itemize}

Only when both the source and target data is clean data, we are witnessing a lower F1-score compared to the F1-score of 0.91/0.92 obtained by CRNNs. 

Looking at the tabulated results in Table \ref{tab:table-2} and Table \ref{tab:yoho9dB}, YOHO outperforms CRNNs on the -9dB SNR VOICe dataset. This is despite the fact that implementation of the YOHO model in our experiment did not involve any adaptation of the model using target data $D_t$, i.e., the model is able to report higher F1-scores on most of the target data without any fine-tuning on the target data involved.

\begin{table}[ht]
\centering
\caption{\label{tab:table-2} Results of CRNN on VOICe from original paper \cite{4_voice} using -9dB SNR data.}
\begin{tabular}{|c|c|c|c|}

\hline
$D_s$ / $D_t$  & $F_{1_S}^{S_{te}}$ & $F_{1_S}^{T_{te}}$ & $F_{1_A}^{T_{te}}$ \\ \hline
\begin{tabular}[c]{@{}c@{}}V. / O.\\ V. / I.\end{tabular}                & \begin{tabular}[c]{@{}c@{}}0.80 ± 0.003\\ 0.81 ± 0.008\end{tabular}                & \begin{tabular}[c]{@{}c@{}}0.73 ± 0.003\\ 0.69 ± 0.002\end{tabular}                & \begin{tabular}[c]{@{}c@{}}0.74 ± 0.006\\ 0.66 ± 0.018\end{tabular}                \\ \hline
\begin{tabular}[c]{@{}c@{}}O. / V.\\ O. / I.\end{tabular}                & \begin{tabular}[c]{@{}c@{}}0.75 ± 0.010\\ 0.75 ± 0.008\end{tabular}                & \begin{tabular}[c]{@{}c@{}}0.77 ± 0.007\\ 0.70 ± 0.011\end{tabular}                & \begin{tabular}[c]{@{}c@{}}0.76 ± 0.023\\ 0.69 ± 0.013\end{tabular}                \\ \hline
\begin{tabular}[c]{@{}c@{}}I. / V.\\ I. / O.\end{tabular}                & \begin{tabular}[c]{@{}c@{}}0.73 ± 0.002\\ 0.74 ± 0.002\end{tabular}                & \begin{tabular}[c]{@{}c@{}}0.75 ± 0.008\\ 0.75 ± 0.008\end{tabular}                & \begin{tabular}[c]{@{}c@{}}0.73 ± 0.007\\ 0.70 ± 0.033\end{tabular}                \\ \hline
\begin{tabular}[c]{@{}c@{}}Ce. / V.\\ Ce. / O.\\ Ce. / I.\end{tabular}   & \begin{tabular}[c]{@{}c@{}}0.92 ± 0.003\\ 0.92 ± 0.003\\ 0.91 ± 0.004\end{tabular} & \begin{tabular}[c]{@{}c@{}}0.50 ± 0.007\\ 0.49 ± 0.022\\ 0.50 ± 0.007\end{tabular} & \begin{tabular}[c]{@{}c@{}}0.43 ± 0.011\\ 0.42 ± 0.010\\ 0.40 ± 0.095\end{tabular} \\ \hline
\end{tabular}
\end{table}


\begin{table}[ht]
\centering
 \centering
    \caption{Results of YOHO on -9dB SNR data.}

\begin{tabular}{|c|c|c|c|c|}
\hline
$D_s / D_t$ & Ce           & V            & O            & I            \\ \hline
Ce                        & 0.86 ± 0.008 & 0.75 ± 0.005 & 0.73 ± 0.011 & 0.72 ± 0.007 \\ \hline
V                         & 0.84 ± 0.007 & 0.83 ± 0.004 & 0.77 ± 0.004 & 0.74 ± 0.009 \\ \hline
O                         & 0.82 ± 0.002 & 0.82 ± 0.007 & 0.80 ± 0.006 & 0.77 ± 0.004 \\ \hline
I                         & 0.81 ± 0.002 & 0.82 ± 0.010 & 0.82 ± 0.002 & 0.80 ± 0.003 \\ \hline
\end{tabular}

  \label{tab:yoho9dB}

\end{table}





\subsection{Hyperparameters}

First, when splitting the audio files into overlapping windows, window length was set to 2.56s and hop/jump length was set to 1.96s. The sampling rate of the audio files being read was 44,100Hz. 

We used the Adam optimizer with a learning rate of 0.001. We also used a batch size of 32, and early-stopping callback \cite{22_earlystopping} to train the network where the validation loss was monitored. If the change in validation loss remained less than 0.1 (min-delta) for 5 epochs (patience), then the training of the model would be stopped. Similar to \cite{3_yoho}, L2 normalization with regularisation parameter set to 0.001 and spatial dropout was also employed. SpecAugment \cite{23_specaugment} was used to augment the model inputs - log-mel spectrograms, by randomly dropping a sequence of frequency bins or time bins from the input. 

Finally, we used the sed\_eval toolbox \cite{24_sed_eval}, to assess the performance of the models on the noisy and clean data and to generate the F1-scores and the error rates, reported in Section 4.




\section{Conclusion}

In this paper, we evaluated the robustness of the YOHO algorithm on noisy data. The output layer of YOHO model was adapted and then evaluated on the VOICe dataset. The results showed that the YOHO model was able to outperform the CRNN model on the same data. On the basis of these results, it would seem that YOHO is a robust algorithm and is able to perform well not only on noisy but also on unseen data. Moreover, the much lower inference times of YOHO compared to CRNNs on other datasets \footnote{CRNN inference times for VOICe dataset were not available. Hence we did not report the inference times of YOHO on the VOICe dataset.} reported in the YOHO paper \cite{3_yoho} showed YOHO is a fast, accurate and robust model for sound event detection. We hope that these results would ins pire future studies where the accuracy and low-inference times of YOHO can potentially be used in portable SED applications.

\bibliographystyle{IEEEtran}
\bibliography{bibliography.bib}

\end{document}